\documentclass[]{mn2e}
\usepackage{psfig}
\usepackage{mnras_cite}
\newcommand{\gal}{{\rm gal}}
\newcommand{\zgal}{{\rm z-gal}}
\newcommand{\lin}{{\rm lin}}
\newcommand{\veck}{{\bf k}}

\newcommand{\om}{\Omega_m}

\newcommand{\ob}{\Omega_B}

\newcommand{\ol}{\Omega_\Lambda}

\newcommand{\on}{\Omega_\nu}

\newcommand{\ns}{n_S}

\newcommand{\As}{A^2_S}

\newcommand{\kfid}{k_{\rm fid}}

\newcommand{\Pinit}{P_{\Phi}}

\newlength{\tskip}
\newlength{\colwidth}
\newcommand{\colskip}{@{\hspace{0.3in}}}

\newcommand{\beq}{\begin{equation}}
\newcommand{\eeq}{\end{equation}}
\newcommand{\beqa}{\begin{eqnarray}}
\newcommand{\eeqa}{\end{eqnarray}}

\begin{document}

\title[Galaxy Power Spectrum and Cosmological Parameters]{Non-linear 
Galaxy Power Spectrum and Cosmological Parameters}
\author[Cooray]{Asantha Cooray\\
Theoretical Astrophysics, Mail Code 130-33, Caltech, Pasadena, CA
  91125, USA}

\maketitle

\begin{abstract}
The galaxy power spectrum is now a well-known tool of precision cosmology.
In addition to the overall shape, 
baryon oscillations and the small-scale suppression of power by massive neutrinos capture complimentary information
on cosmological parameters when compared to the angular power spectrum of cosmic microwave background anisotropies. 
We study both the real space and redshift space galaxy power spectra 
in the context of non-linear effects and model them  based on the halo approach to large scale
structure clustering. We consider potential systematic in the cosmological parameter
determination when non-linear effects are ignored and the galaxy power spectrum is described with
the linear power spectrum scaled by a constant bias factor. 
We suggest that significant improvements can be made when non-linear effects are taken into account 
as a power-law contribution  with two additional parameters to be determined from the data. In addition to
cosmological parameters through galaxy clustering, such an approach allow a determination of useful information
related to astrophysics on how galaxies occupy dark matter halos.
\end{abstract}


\section{Introduction}

Through a statistical measurement of information related to inhomogeneities in the universe,
the galaxy power spectrum provides a strong tool in the present era of precision cosmology. 
In addition to information related to the primordial power spectrum, 
characterized by a slope and a normalization, through 
a turnover at $k_{\rm eq} \equiv \sqrt{2\Omega_mH_0^2(1+z_{\rm eq})}$,
the overall shape of the galaxy power spectrum captures cosmological
information related to the horizon at the matter-radiation equality. 
In addition to the overall shape, additional features in the galaxy power spectrum allow improved measurements of
several cosmological parameters. Similar to the oscillatory features in the CMB anisotropy power spectrum, the matter power spectrum is expected to exhibit the presence of baryons through oscillations. These features are associated with the sound
horizon at the end of the Compton-drag epoch  and capture information related to $\Omega_mh^2$ and $\Omega_bh^2$ \cite{EisHu98}.

Unlike well-known oscillatory features in the angular power spectrum of cosmic microwave background (CMB) 
anisotropies, baryon related oscillations in the
matter power spectrum of the large-scale structure are highly suppressed due to the low baryon-to-dark matter ratio. 
For currently favored $\Lambda$CDM cosmologies, baryon oscillations in the matter power spectrum 
have amplitude variations at the level of $\sim$ 8\% with effective widths in Fourier space of order $\Delta k \sim 0.03$ h Mpc$^{-1}$.
While the ideal way to detect baryon oscillations is to measure the three-dimensional dark matter power spectrum directly,
unfortunately, there are no useful probes of this quantity. The effects such as the weak gravitational lensing,  which trace 
matter fluctuations, only provide information related to the projected power spectrum over a  broad window function in the
redshift space (e.g., Bartlemann \& Schneider 2001).  This leads to  an  averaging of  small features such as baryon oscillations
such that  they remain undetectable even considering most favorable scenarios.

Galaxy redshift surveys, on the other hand, allow a measurement of the three dimensional power spectrum of the galaxy distribution.
Since galaxies are expected to trace matter fluctuations, at least in large and linear scales, the galaxy power spectrum has been
pursued for an  observational detection of baryon oscillations. When detected, baryon oscillations are expected
to allow significant improvements in parameter determination by breaking various degeneracies associated with
the interpretation of CMB data  \cite{Eisetal98,BlaGla03,Eis02,Coo02a}.
While expectations for precision parameter measurements are generally high, these have been tested to some extent with
measured power spectra from redshift surveys such as the 2dF Galaxy Redshift Survey 
(2dFGRS; Colless et al. 2001)  
when combined with recent results from CMB anisotropy experiments (e.g., Percival et al. 2002).
The same surveys have allowed first attempts to detect baryon oscillations, though,  
there is still no significant evidence for them (e.g., Percival et al. 2001; also, see, Miller 
Nichol \& Batuski 2001). This can be understood by the fact that
three-dimensional redshift surveys, so far, lack the required volume, with sizes of
order $L \sim 2\pi/\Delta k \sim 300$ h$^{-1}$ Mpc in each dimension, 
to reliably resolve them.

In addition to galaxy redshift surveys, a number of additional observational efforts are either under way or planned 
to image the large-scale structure out 
to a redshift of a few.  These wide-field surveys typically cover tens to thousands of square degrees on the sky and include 
weak gravitational lensing shear observations with instruments such as the Supernova Acceleration Probe (SNAP; Massey et al. 2003) 
and the Large Aperture Synoptic Survey Telescope 
(LSST; Tyson, Wittmann \& Angel 2000), and observations of the Sunyaev-Zel'dovich effect \cite{SunZel80} with 
dedicated small angular scale CMB telescopes such as the South Pole Telescope (SPT; Stark et al. 1998).
These surveys are expected to produce catalogs of dark matter halos, which in the
case of lensing and SZ surveys are expected to be essentially
mass selected \cite{Witetal01,Holetal00}.  While the halo number count as a function of redshift
is a well-known cosmological test \cite{Haietal01},  one can also consider the additional
information supplied by the clustering of halos. In particular, similar to the approach in galaxy redshift surveys,
one can measure the power spectrum of clustering associated with these halos, as a whole, and use that power spectrum for
cosmological studies. In addition to the shape, cosmological information comes from the redshift evolution of {\it rulers}
that can be calibrated through CMB data \cite{Cooetal01,BlaGla03}.

The cosmological studies based on the galaxy power spectrum and related halo clustering information, unfortunately, is affected by
non-linearities. While sources are expected to trace 
inhomogeneities in mass, due to the non-linear evolution of gravitational perturbations at late times, the
galaxy power spectrum at small scales departs significantly from the linear description corrected for source bias.
The enhancement of power due to non-linearities erases oscillatory features \cite{Meietal99}, while effects such as 
redshift-space distortions further contribute to their disappearance.
Due to the cosmological significance of baryon oscillations, suggestions have  already been made to detect them at
redshifts of a few since the non-linear scale at high redshift is expected to move to smaller scales than today \cite{Eis02,BlaGla03}. 

In the case of current measurements, the cosmological interpretation is restricted only for 
large scales where clustering is expected to be linear. The standard approach to describe the
galaxy power spectrum involves the linear power spectrum scaled by a constant bias factor 
(e.g., Percival et al. 2001). While  slight modifications to the bias description have been considered 
\cite{ElgLah03}, there is still no guidance as to
how non-linearities affect cosmological parameter measurements and what approaches can be made to
improve such estimates. Additionally, by restricting cosmological interpretation to specific range of
wave numbers, it is likely that we have only extracted limited information from current data.
For example, in the case of the 2dFGRS, analyzes have only been considered out to a wave number of 0.15 h Mpc$^{-1}$,
while measurements span to considerably smaller scales.  It is clear that further studies are needed on
galaxy clustering well in to the non-linear regime and to understand what information can be
extracted from the non-linear regime of clustering, or at least, from transition regime where non-linearities
become important.

The purpose of this paper is to understand the onset of non-linearities in the galaxy power spectrum. While detailed numerical
simulations are useful for this purpose, here, we make use of the halo approach to galaxy clustering 
\cite{Sel00,PeaSmi00,Scoetal01,SheDia01,BerWei02}.
The halo approach provides a useful tool to understand certain aspects of clustering and is physically consistent model
without any prior assumptions on the biasing nature of galaxies with respect to the dark matter distribution (for a review,
see, Cooray \& Sheth 2002).
We study how cosmological parameter measurements can be improved by extracting information from mildly non-linear scales 
that are  currently ignored. In particular, we pay attention to systematics that are introduced to cosmological
parameter measurements when the galaxy power spectrum is modeled simply with a scaled version of the linear power spectrum.
We show that parameter estimates are  biased away from the true values and that these departures can be significantly reduced when the non-linear contribution, at least near the regime of transition from linear to non-linear clustering, is 
modeled as a power-law.Since parameters related to this description is determined from data simultaneously with cosmological parameter estimates, this
approach leads to a slight degradation of errors associated with cosmological
parameters alone. The main advantage is that these estimates, however, remain either unbiased or biased at a level that is negligible.  
The suggested approach can be easily implemented with current studies and when interpreting results related to
completed 2dFGRS and SDSS surveys \cite{Yoretal00}, among others. 
The suggested approach also avoids an arbitrary distinction as to what scales correspond to  the linear regime of
clustering while also allowing the maximum extraction of cosmological information from data. 

The paper is organized as following. In the next Section, we outline main considerations related to 
the halo approach to galaxy clustering. We refer the reader to review by Cooray \& Sheth (2003) for full details of
this approach. We make use of the Fisher matrix approach to quantify biases and expected errors on parameter measurements in Section~3. In Section~4, we conclude with a summary.

\begin{figure*}
\centerline{\psfig{file=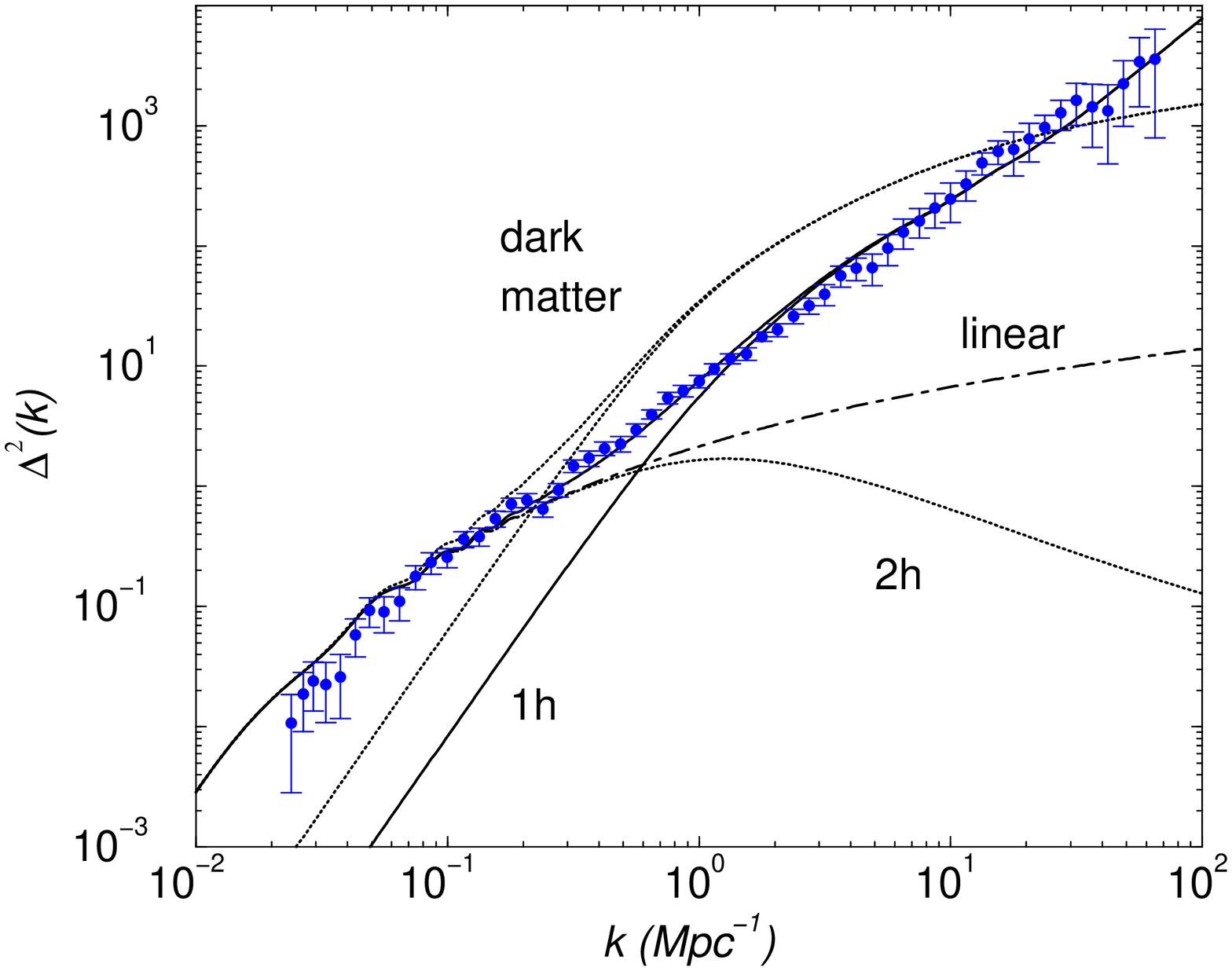,width=110mm,angle=-90}}
\centerline{\psfig{file=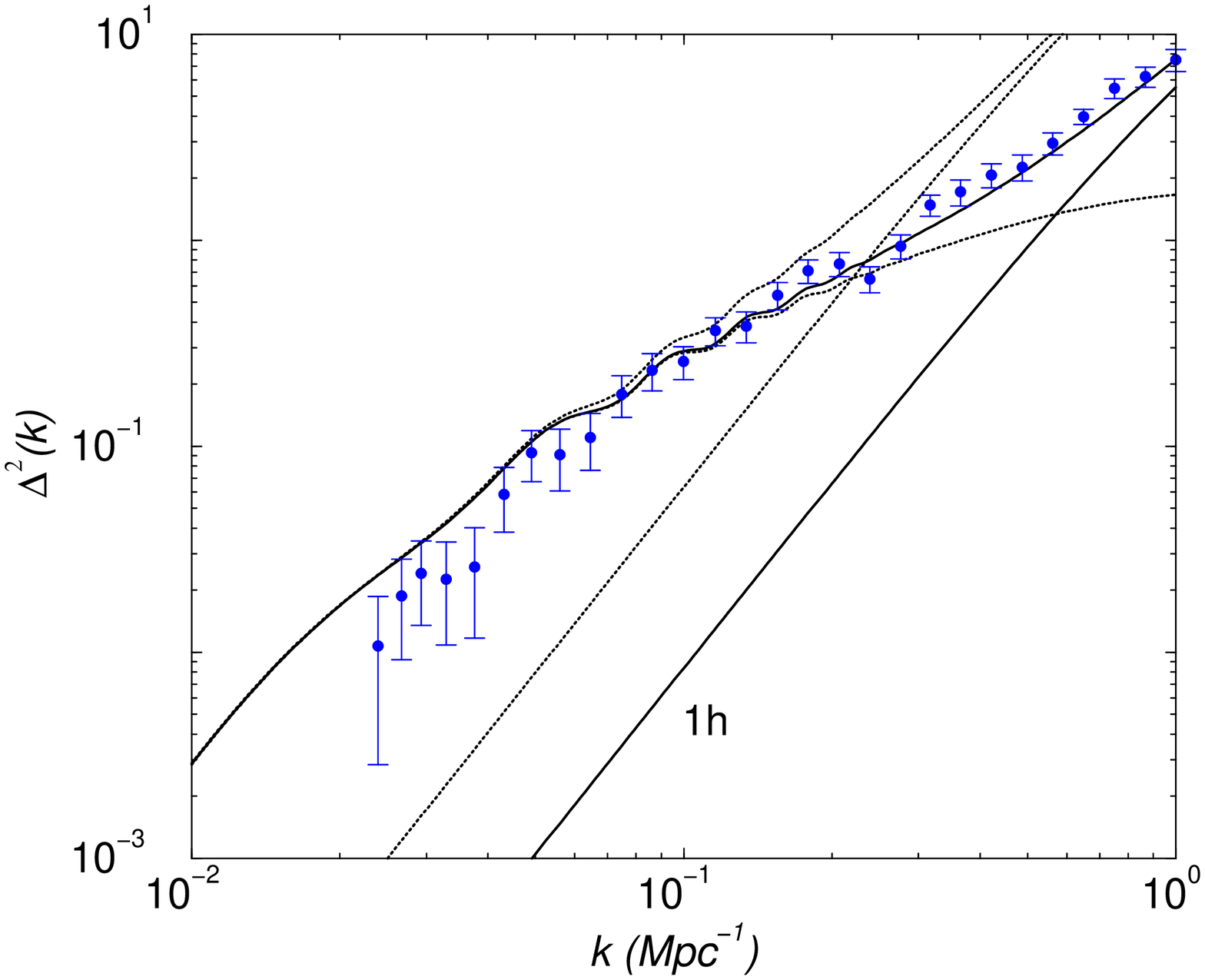,width=110mm,angle=-90}}
\caption{The galaxy power spectrum in real space (solid lines) and a comparison to the clustering in dark matter (dotted lines). 
We show the 1- and 2-halo contributions to the galaxy power spectrum and the 1-halo contribution to the dark matter power spectrum.
For reference, the dot-dashed line shows the linear power spectrum. The measured data points
represent the power spectrum of the IRAS Point Source Catalog Redshift Survey (PCSz; Saunders et al. 2000) as measured 
by Hamilton \& Tegmark (2002). In the bottom panel, we consider the transition regime between linear to non-linear clustering.
As we discuss later, the transition to non-linear clustering can be described via a combination of the linear power spectrum scaled
with a constant, scale-free, bias and an additional contribution which is a power-law.}
\label{fig:power}
\end{figure*}

\section{Halo Approach to Clustering}

\subsection{Real Space Clustering}

The halo approach provides a physically motivated scheme to understand non-linear clustering of dark matter and various
tracer populations of the large scale structure (see Cooray \& Sheth 2002 for more details).
Under this description, the galaxy spectrum takes the form of
\begin{eqnarray}
 P_\gal(k) &=& P^{1h}_{\gal}(k) + P^{2h}_{\gal}(k)\, ,
               \qquad\qquad{\rm where}\nonumber\\
 P^{1h}_{\gal}(k) &=& \int dm \, n(m) \,
                    \frac{\left< N_\gal(N_\gal-1)|m\right>}{\bar{n}_\gal^2}\,
                    |u_\gal(k|m)|^p \,,\nonumber\\
 P^{2h}_{\gal}(k) &=& P^\lin(k) \left[ \int dm\, n(m)\, b_1(m)\,
 \frac{\left< N_\gal|m\right>}{\bar{n}_\gal}\, u_\gal(k|m)\right]^2 . \nonumber \\
 \label{eqn:gal}
\end{eqnarray}
Here,
\begin{equation}
 \bar{n}_\gal = \int dm \, n(m)\, \left< N_\gal|m \right> \,  ,
 \label{eqn:barngal}
\end{equation}
denotes the mean number density of galaxies
and
\begin{equation}
 u_{\rm gal}(k|m) = \int_0^{r_{vir}} dr\ 4\pi r^2\,{\sin kr\over kr}\
{\rho_{\rm gal}(r|m)\over m} \, ,
\label{eqn:yint}
\end{equation}
denotes the normalized Fourier transform of the galaxy density distribution within
a halo of mass $m$. 
The standard assumption here is that galaxies trace dark matter within halos such that
one can utilize the dark matter distribution given by analytic forms such as the NFW \cite{Navetal96} profile.
An improved approximation will be to use the density distribution defined by sub-halos.
Since relevant profiles related to substructure is still not well studied numerically, we make use of
the dark matter density profile to describe the galaxy distribution within halos.
In equations~\ref{eqn:gal} and \ref{eqn:barngal}, $n(m)$ is the halo mass
function (e.g., Press \& Schechter 1974) and $b_1(m)$ is the halo bias relative to linear clustering \cite{MoWhi96}.

An important ingredient in the halo description of galaxy clustering is the
halo occupation number \cite{Kauetal99,Benetal00,Beretal03}.
Following  Sheth \& Diaferio (2001), for the illustration of non-linear effects related to galaxy clustering,
we take a description motivated by semi-analytic models:
\begin{eqnarray}
\langle N_\gal|m \rangle = \left\{\begin{array}{ll}
N_0  & 10^{11}\,M_{\sun} h^{-1}  \leq m\leq M_{B}\\
  N_0\,(m/M_{\rm B})^{\alpha} & m>M_{\rm B}  
\end{array}\right.
\label{eqn:galcounts}
\end{eqnarray}
where  we set $M_{\rm B} = 4\times 10^{12}\,M_\odot/h$ and take $\alpha$ and $N_0$ as free parameters.
In our fiducial description, these two parameters take numerical values of 0.8 and 0.7, respectively.

In addition to the mean number of galaxies as a function of halo mass, 
for the one-halo term of the power spectrum (equation~\ref{eqn:gal}), we need
information related to the second moment of the halo occupation number, $\left< N_\gal(N_\gal-1)|m\right>$.
If the probability distribution function for galaxy occupation is Poisson, 
then  $\langle N_\gal(N_\gal -1)|m \rangle = \langle N_\gal|m \rangle^2$,
though departures from the Poisson description is clearly visible in numerical data \cite{Beretal03}.
Following Scoccimarro et al. (2001), we make use of the binomial distribution, matched to numerical data,
 to obtain a convenient approximation. The second moment is then
\begin{equation}
 \langle N_\gal(N_\gal -1)\rangle^{1/2} =
 \beta(m)\,\langle N_\gal|m\rangle \, ,
\label{binomial}
\end{equation}
where $\beta(m) = \log \sqrt{m/10^{11} h^{-1} M_{\sun}}$ for
$m < 10^{13} h^{-1} M_{\sun}$ and $\beta(m)=1$ thereafter.
Note that in equation~\ref{eqn:gal}, 
the simplest approach is to set $p=2$ when calculating $P^{1h}_{dm}(k)$.  
In halos which contain only a single galaxy, however, it is natural to assume that this galaxy sits at the center.
Thus, an improved model for the galaxy distribution is to take $p=2$ when
$\left< N_\gal(N_\gal-1)\right>$ is greater than unity and $p=1$ otherwise.

In Fig.~1, we show the galaxy power spectrum calculated following the halo model and a comparison 
to the non-linear dark matter power spectrum, again calculated under the halo model; to describe dark matter clustering,
one removes information related to the halo occupation number and the number density of galaxies from equation~\ref{eqn:gal}.
Note that in Fig.~1 and throughout, we plot the logarithmic power spectrum, $\Delta^2(k) \equiv k^3P(k)/2\pi^2$ instead of
$P(k)$ itself.
We also show a measurement of the real space power spectrum from the IRAS PSCZ survey \cite{Sauetal90} as determined by 
Hamilton \& Tegmark (2002).
In the bottom plot of Fig.~1, we show the transition regime between linear and non-linear clustering.
At large scales, where the two-halo term dominates,
the galaxy power spectrum traces the linear clustering with a constant bias. One can understand this by noting that
at large scales, $u_\gal(k|m)\to 1$ and the galaxy power spectrum simplifies to
\begin{equation}
 P_{\gal}(k) \approx b_\gal^2\, P^\lin(k),
\end{equation}
with
\begin{equation}
 b_\gal = \int dm\, n(m)\, b_1(m)\,
          \frac{\left< N_\gal|m\right>}{\bar{n}_\gal} 
\end{equation}
denoting the mean bias factor of the galaxy population.

This is consistent with the standard assumption in the literature that the
galaxy bias, at large scales, is constant. The drawback with this description is that the
physical scale below which this assumption breaks down is ill defined.
It is generally assumed that the on set of non-linearity is at wave numbers $\sim$ 0.1 h Mpc$^{-1}$, though this is
only based on numerical simulations  of the non-linear dark matter power spectrum.
While such a scale is valid for the dark matter distribution, when considering galaxies,  there is no  reason
why the same situation should hold. In the case of the halo approach to galaxy clustering, 
the onset of non-linearities, by which we assume the dominance of the 1-halo term, is model dependent.
To illustrate this, in Fig.~2, we model the galaxy power spectrum as a function of $\alpha$ and $N_0$. 
Note that the transition is strongly sensitive to $\alpha$ such that
when $\alpha$ is low, the galaxy power spectrum traces the linear clustering down to relatively smaller scales, $\sim$ 0.5 Mpc$^{-1}$,
while when $\alpha$ is high ($>$ 0.8), the onset of non-linearity is near 0.1 Mpc$^{-1}$.
If $\langle N_\gal|m \rangle$ simply scales as $N_0(m/M_b)^\alpha$, then the galaxy power spectrum is
independent of $N_0$, as one renormalizes by the galaxy number density which is also an integral of
this quantity weighted by the mass function. In Fig.~2, we see slight dependences on $N_0$ since the halo occupation number
was modeled with two parts, one which is a constant and another involving a power-law. In this case,
the 1-halo term is slightly dependent on the numerical value of $N_0$.

\begin{table*}
\begin{center}
\caption{Marginalized Errors}
\end{center}
\begin{center}
\begin{tabular}{lccc \colskip ccc\colskip ccc}
\hline \hline
& \multicolumn{3}{c\colskip}{k $<$ 0.1 Mpc$^{-1}$} & \multicolumn{3}{c}{k $<$ 0.2 Mpc$^{-1}$} & \multicolumn{3}{c}{k $<$ 0.5 Mpc$^{-1}$} \\
Parameter & Error & Bias & $|$Bias$|$/Error & Error & Bias & $|$Bias$|$/Error & Error &Bias & $|$Bias$|$/Error \\
\hline
$\ln(\ob h^2)$                    & 1.35 & 0.15 & 0.11  & 0.22 & 0.15 & 0.73 &  0.20 & 0.29 & 1.44 \\
$\ln(\om h^2)$                    & 1.29 & 0.06 & 0.05  & 0.31 & 0.05 & 0.17 &  0.29 & 0.08 & 1.75\\
$m_\nu(eV) \propto \on h^2$       & 5.64 & -1.01 & 0.18 & 1.59 & -1.90 & 1.19&  1.35 & -7.54 & 5.56 \\
$\ol$                             & $\infty$ & - & -   & $\infty$ & - &   -   &  $\infty$ & - & - \\
$w$                               & $\infty$ & - & -   & $\infty$ & - &   -   & $\infty$ & - & - \\
$\ns(\kfid)$                      & 0.58 & -0.04 & 0.06 & 0.09 & -0.22& 2.34 & 0.08 & -0.66  & 8.74 \\
$\ln\Pinit(\kfid)\equiv\ln\As$    & $\infty$ & - & -      & $\infty$ & -    & -  & $\infty$ & - & - \\
$\ln b$                           & $\infty$ & - & -      & $\infty$ & -    &   -& $\infty$ & - & - \\
\hline
$\ln(\ob h^2)$                    & 0.03 & 0.00 & 0.01  & 0.02 & 0.02  & 0.76 &  0.02 & 0.23 & 10.56 \\
$\ln(\om h^2)$                    & 0.06 & 0.00 & 0.03  & 0.05 & 0.06  & 1.25 &  0.04 & 0.39 & 8.38 \\
$m_\nu(eV) \propto \on h^2$       & 0.38 & -0.06 & 0.09 & 0.23 & -0.94 & 4.23 &  0.21 & -3.48 & 16.77 \\
$\ol$                             & 0.07 & 0.00 & 0.00   & 0.06 & 0.02 & 0.33 &  0.06 &0.16 & 2.92 \\
$w$                               & 0.29 & -0.00 & 0.03   & 0.27 &  -0.3 & 1.11 & 0.26 & -2.04 & 7.85 \\
$\ns(\kfid)$                      & 0.018 &-0.00 & 0.04 & 0.015 & -0.03& 2.02& 0.01 &-0.24 & 18.67 \\
$\ln\Pinit(\kfid)\equiv\ln\As$    & 0.021 & 0.00 & 0.00      & 0.017 & 0.02& 1.07& 0.016 & 0.19 & 12.00 \\
$\ln b$                           & 0.040 & 0.00 & 0.09      & 0.034 & 0.06 & 1.93& 0.033 & 0.06 & 1.77 \\
\hline
$\ln(\ob h^2)$                    & 0.005 & 0.00 & 0.03  & 0.005 & 0.01 & 1.17 &  0.005 & 0.04 & 6.88 \\
$\ln(\om h^2)$                    & 0.015 & 0.01 & 0.06  & 0.013 & 0.04 & 2.84 &  0.012 & 0.20 & 16.21 \\
$m_\nu(eV) \propto \on h^2$       & 0.17  & -0.02 & 0.10 & 0.12 & -0.72 &  5.78 & 0.09 & -3.47 & 38.43 \\
$\ol$                             & 0.05 & 0.00 & 0.00   & 0.049 & 0.00 & 0.03   &  0.05 & 0.022 & 0.44 \\
$w$                               & 0.15 & -0.00 & 0.03   & 0.17 &  -0.15&  0.89   & 0.17 & -0.68 & 4.01 \\
$\ns(\kfid)$                      & 0.004 & -0.00 & 0.04 & 0.004 & -0.01 & 2.43 & 0.003 & -0.05 & 16.95 \\
$\ln\Pinit(\kfid)\equiv\ln\As$    & 0.005 & 0.00 & 0.01     & 0.005 & 0.01  & 1.03  & 0.004 & 0.04 & 8.92 \\
$\ln b$                           & 0.017 & 0.00 & 0.06      & 0.016 & 0.05    & 2.92  & 0.013 & 0.36 & 26.91 \\
\hline
\end{tabular}
\end{center}
NOTES.---%
Parameter error calculation assume following fiducial model:
$\om=0.3$, $\ob=0.04$, $\ol=0.7$, $h=0.72$, $w=-1$ and $\on=0$. We fix other interesting parameters related to CMB, 
$\tau=0.1$ and $T/S=0$, and assume a flat-cosmology with $\Omega_K=0$.  We quote  $1\sigma$ errors.
The top, central, and bottom divisions represent the parameter errors from galaxy clustering, galaxy clustering with
the full 3 year WMAP data, and galaxy clustering with Planck data.
A parameter error of infinity implies that the parameter affects observables but
is not constrained due to degeneracies; in particular, the normalization and bias are all degenerate and is only broken
through the addition of CMB data.
\end{table*}

\begin{table*}
\begin{center}
\caption{Marginalized Errors With non-Linearities Accounted by a Power-Law}
\end{center}
\begin{center}
\begin{tabular}{lccc \colskip ccc\colskip ccc}
\hline \hline
& \multicolumn{3}{c\colskip}{k $<$ 0.1 Mpc$^{-1}$} & \multicolumn{3}{c}{k $<$ 0.2 Mpc$^{-1}$} & \multicolumn{3}{c}{k $<$ 0.5 Mpc$^{-1}$} \\
Parameter & Error & Bias & $|$Bias$|$/Error & Error & Bias & $|$Bias$|$/Error & Error &Bias & $|$Bias$|$/Error \\
\hline
$\ln(\ob h^2)$                    & 7.33 & 0.42 & 0.05  & 0.42 & 0.03 & 0.07   & 0.20 & 0.005 & 0.02 \\
$\ln(\om h^2)$                    & 4.88 & 0.24 & 0.05  & 0.38 & 0.03 & 0.08   &  0.33 & 0.02 & 0.06\\
$m_\nu(eV) \propto \on h^2$       & 7.27 & -0.14 & 0.02 & 1.74 & -0.10 & 0.06  &   1.50 & -0.04 & 0.03 \\
$\ol$                             & $\infty$ & - & -   & $\infty$ & - &   -      &  $\infty$ & - & - \\
$w$                               & $\infty$ & - & -   & $\infty$ & - &   -    & $\infty$ & - & - \\
$\ns(\kfid)$                      & 0.78 & -0.02 & 0.03 & 0.12 & -0.01& 0.08   & 0.11 & -0.01  & 0.09 \\
$\ln\Pinit(\kfid)\equiv\ln\As$    & $\infty$ & - & -      & 2.36 & -0.11& 0.06  & 0.94 & 0.09 & 0.09 \\
$\ln b$                           & $\infty$ & - & -      & 1.64 & 0.11   & 0.06 &0.82 & 0.07 & 0.08 \\
$A_{\rm nl}$                  & $\infty$ & - & -      & 3.95 & 0.31   & 0.08 & 0.41 & 0.06 & 0.14 \\
$\alpha_{\rm nl}$                 & $\infty$ & - & -      & 3.58 & -0.14  & 0.04  & 0.45 & -0.05 & 0.11 \\
\hline
$\ln(\ob h^2)$                    & 0.25 & 0.00 & 0.02  & 0.024 & 0.00  & 0.05 &  0.02 & 0.00 & 0.05 \\
$\ln(\om h^2)$                    & 0.06 & 0.00 & 0.04  & 0.05 & 0.00  & 0.02 &  0.05 & 0.00 & 0.04 \\
$m_\nu(eV) \propto \on h^2$       & 0.41 & -0.03 & 0.07 & 0.25 & -0.01 & 0.04 &  0.23 & -0.01 & 0.04 \\
$\ol$                             & 0.06 & 0.00 & 0.01   & 0.06 & 0.00 & 0.02&  0.06 & 0.00 & 0.02 \\
$w$                               & 0.30 & 0.01 & 0.04   & 0.28 &  0.02 & 0.06& 0.27 & 0.00 & 0.02 \\
$\ns(\kfid)$                      & 0.019 & 0.00 & 0.04 & 0.016 & -0.00& 0.07& 0.01 & -0.00 & 0.01 \\
$\ln\Pinit(\kfid)\equiv\ln\As$    & 0.020 & 0.00 & 0.03      & 0.018 & 0.00& 0.08& 0.016 & 0.00 & 0.05 \\
$\ln b$                           & 0.13 & 0.00 & 0.06      & 0.036 & 0.00 & 0.03& 0.034 & 0.00 & 0.06 \\
$A_{\rm nl}$                  & $\infty$ & - & -      & 2.27 & 0.21 & 0.09  & 0.22 & 0.04 & 0.18 \\
$\alpha_{\rm nl}$                 & $\infty$ & - & -      & 1.98 & -0.12& 0.06&   0.23 & 0.03 & 0.13  \\
\hline
$\ln(\ob h^2)$                    & 0.005 & 0.00 & 0.00  & 0.006 & 0.00 & 0.01 &  0.005 & 0.00 & 0.04 \\
$\ln(\om h^2)$                    & 0.015 & 0.00 & 0.02  & 0.014 & 0.00 & 0.02 &  0.014 & 0.00 & 0.07 \\
$m_\nu(eV) \propto \on h^2$       & 0.18 & -0.00 & 0.02 & 0.14 & -0.00 & 0.01 &  0.10 & -0.00 & 0.01 \\
$\ol$                             & 0.05 & 0.00 & 0.00   & 0.05 & 0.00 & 0.00   &  0.05 &0.00  & 0.00 \\
$w$                               & 0.17 & 0.00 & 0.01   & 0.17 &  0.00&  0.01   & 0.17 & 0.01 & 0.05 \\
$\ns(\kfid)$                      & 0.004 & -0.00 & 0.01 & 0.004 & -0.00 & 0.01 & 0.003 & -0.00 & 0.01 \\
$\ln\Pinit(\kfid)\equiv\ln\As$    & 0.005 & -0.00 & 0.01      & 0.005 & 0.00  & 0.02  & 0.004 & 0.00 & 0.02 \\
$\ln b$                           & 0.12 & 0.01 & 0.08      & 0.025 & 0.00    & 0.08  & 0.016 & 0.00 & 0.06 \\
$A_{\rm nl}$                  & $\infty$ & - & -      & 1.76 & 0.18    & 0.10  & 0.18 & 0.02 & 0.11 \\
$\alpha_{\rm nl}$                 & $\infty$ & - & -      & 1.51 & -0.11    &  0.07& 0.19 & 0.02 & 0.10 \\
\hline
\end{tabular}
\end{center}
NOTES.---%
Same as Table 1. We add two extra parameters: the normalization of the non-linear contribution to
galaxy power spectrum, $A_{\rm nl}$ and its power-law slope, $\alpha_{\rm nl}$, with scale; In the case of our fiducial model and a logarithmic power spectrum, the two parameters have numerical values of 65.7 and 3.04, respectively.
Note the reduction in parameter biases compared to Table~1.
\end{table*}

\begin{figure*}
\centerline{\psfig{file=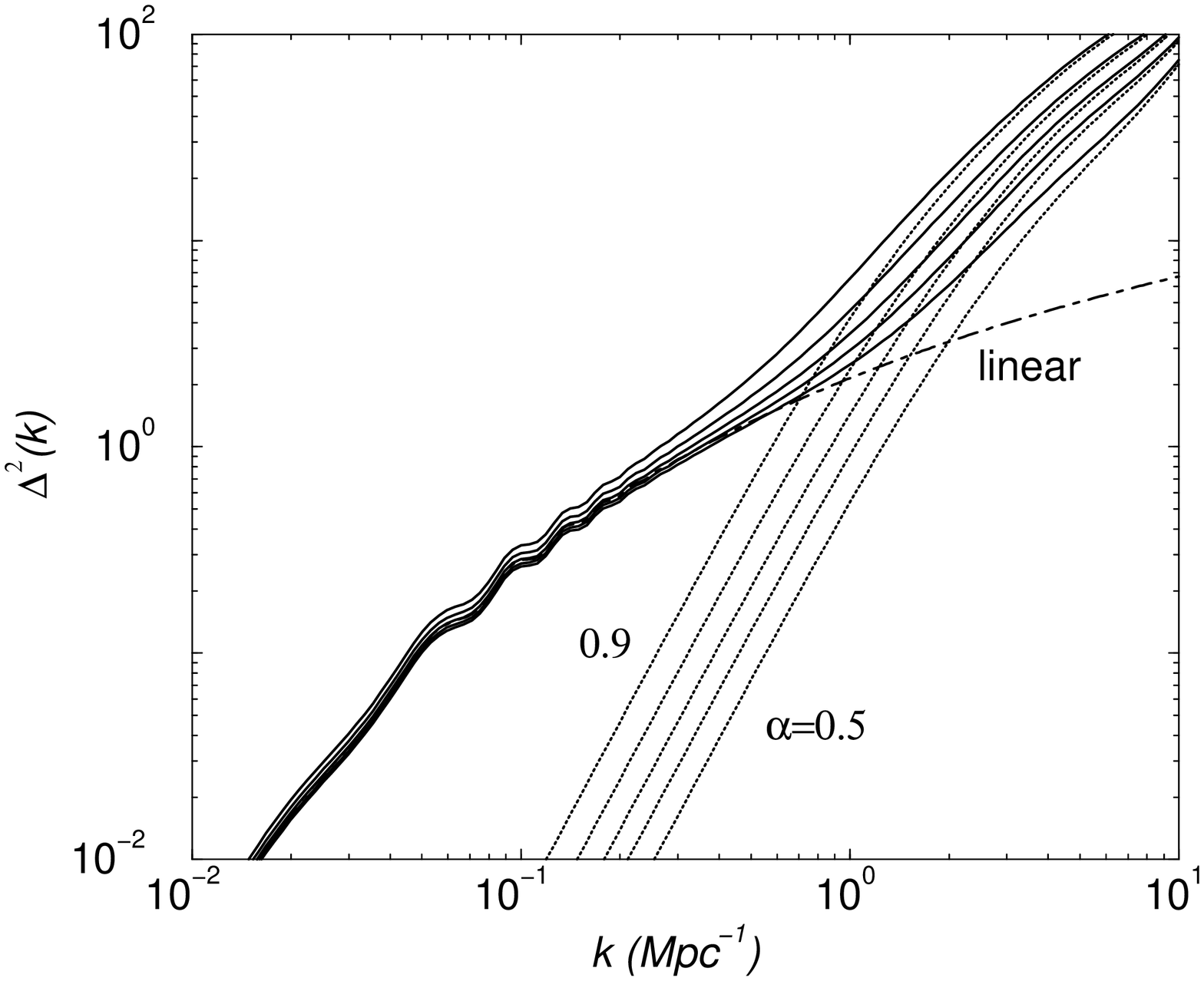,width=110mm,angle=-90}}
\centerline{\psfig{file=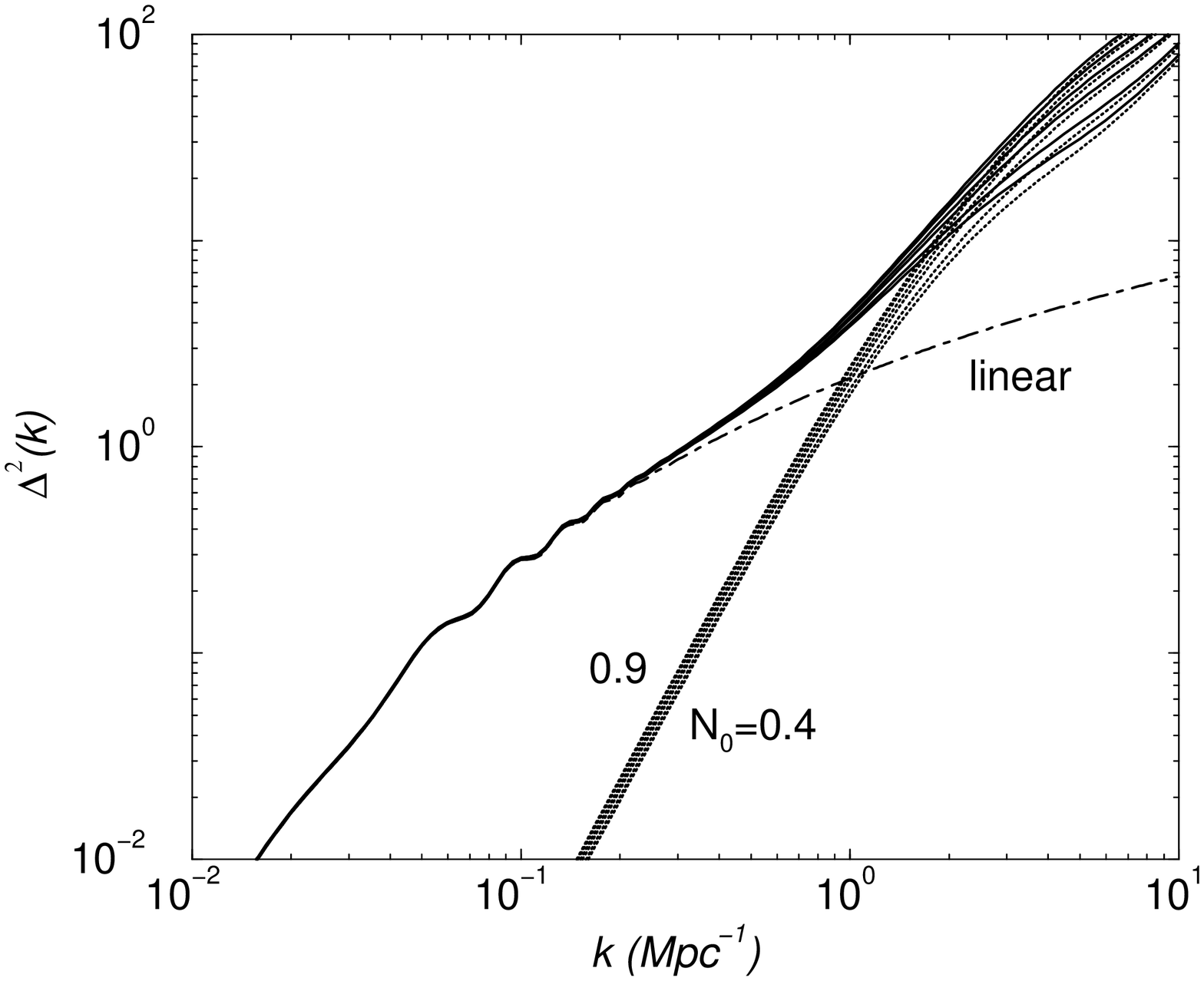,width=110mm,angle=-90}}
\caption{The dependence on the real space galaxy power spectrum on model parameters related to the halo occupation number.
In the top panel, we show variations related to $\alpha$ (with $N_0=0.7$) and the bottom panel shows variations related to
 $N_0$ (with $\alpha=0.8$).
As shown, the slope of the mean number of galaxy as a function of halo mass strongly determines the scale at which
1-halo term, the non-linear contribution, dominates the total power spectrum.}
\label{fig:alpha}
\end{figure*}

\subsection{From Real-Space to Redshift-Space}

So far, we have only considered the real space power spectrum of galaxies. Observationally, 
for example in the case of the 2dF survey, the galaxy power spectrum
is measured from redshift surveys and additional corrections must be taken into account when modeling
clustering in the redshift space.
Following Kaiser (1987), we can write the redshift space fluctuation, $\delta_g^z$, of galaxy density field, relative to
real space fluctuation, $\delta_g$, as
\begin{equation}
\delta_g^z(\veck) = \delta_g(\veck)+\delta_v \mu^2 \,,
\end{equation}
where $\delta_v$ is the divergence of the velocity field and $\mu = \hat{{\bf r}} \cdot \hat{\bf k}$ is the line of sight
angle. At linear scales, 
one can simplify the relation by noting that $\delta_g(\veck) = b_g \delta(\veck)$ and
$\delta_v = f(\Omega_m) \delta(\veck)$ with $f(\Omega_m)=d \log \delta/d \log a \approx \Omega_m^{0.6}$ to obtain
\begin{equation}
\delta_g^z(\veck) = \delta_g(\veck)[1+\beta\mu^2]\, ,
\label{eqn:deltaz}
\end{equation}
where $\beta = f(\Omega_m)/b_g$. The latter is generally measured from redshift surveys
since it allows a determination of $\Omega_m$ through constraints on the bias factor \cite{StrWil95,Outetal01,Peaetal01}.
At linear scales, the distortions increase power by a factor $(1+2/3\beta+1/5\beta^2)$, which when $b_g=1$ is 1.41
for $\Omega_m=0.35$. 
At non-linear scales, virial velocities within halos modify clustering properties. With a Gaussian description for one dimensional 
virial motions, we write
\begin{equation}
\delta_g^z(\veck) = \delta_g e^{-(k\sigma \mu)^2/2} \, .
\end{equation}
Putting all terms together, under the halo approach,
we can write the power spectrum in redshift space as \cite{Sel01}
\begin{eqnarray}
&& P_\zgal(k) = P^{1h}_{\zgal}(k) + P^{2h}_{\zgal}(k)
               \qquad\qquad{\rm where}\nonumber\\
&& P^{1h}_{\zgal}(k) = \nonumber \\
&&\int dm \, n(m) \,
                    \frac{\left< N_\gal(N_\gal-1)|m\right>}{\bar{n}_\gal^2}\,
                   R_p(k\sigma) |u_\gal(k|m)|^p \,,\nonumber\\
&&P^{2h}_{\zgal}(k) = \left(F_g^2+\frac{2}{3}F_vF_g+\frac{1}{5}F_v^2\right)P^\lin(k) \, , \nonumber \\
\label{eqn:galz}
\end{eqnarray}
with
\begin{eqnarray}
F_g &=& \int dm\, n(m)\, b_1(m)\,
\frac{\left< N_\gal|m\right>}{\bar{n}_\gal}\, R_1(k\sigma) u_\gal(k|m) \, , \nonumber \\
F_v &=& f(\Omega_m) \int dm\, n(m)\, b_1(m)\,
R_1(k\sigma) u(k|m)  \, ,
\end{eqnarray}
and
\begin{equation}
R_p(\alpha=k\sigma \sqrt{p/2}) = \frac{\sqrt{\pi}}{2} \frac{{\rm erf}(\alpha)}{\alpha} \, ,
\end{equation}
for $p=1,2$. In equation~(\ref{eqn:galz}), $\bar{n}_\gal$ denotes the mean number density of galaxies 
(equation~\ref{eqn:barngal}).

In order to model velocities within virialized halos,
we make use of a description based on isothermal spheres and in
reasonable agreement with measurements of virial velocities within halos in numerical simulations \cite{SheDia01}.
If $\sigma_{vir}$ denotes the one dimensional velocity dispersion of particles within a halo, then
\begin{equation}
 \sigma_{vir} = \gamma \left[{Gm\over 2 r_{vir}}\right]^{0.5} \, ,
\end{equation}
where the virial radius is related to the halo mass through $M_{vir}=4\pi \Delta \bar{\rho} r_{vir}^3/3$ and $\gamma$
is a free parameter that we use here to demonstrate how uncertainties related to virial motion affect the
predicted non-linear contribution to the redshift-space power spectrum of galaxies.

Even though peculiar velocities increase power at large scales, virial motions within halos lead to a suppression of power. 
In figure~3,  we show the power spectra of dark matter and galaxies in redshift space with a comparison to
to that of the real space. While clustering power is increased at large scales, both for dark matter and galaxies,
the power is reduced substantially at scales where the 1-halo term of the power spectrum dominates. This is easily understandable since
the virial motions within halos lead to decrease in a power when averaged over all angles.

As a comparison to observable data so far, to model the 2dFGRS redshift space power spectrum, 
we move from theory space description to the observable space
by convolving the theory power spectrum over the 2dF survey window function:
\begin{equation}
W_{\rm 2dF}^2(k) = \frac{1}{1+(k/a)^2 + (k/b)^4}
\end{equation}
where $a=0.00342$ and $b=0.00983$ \cite{Elgetal02}. The observed power spectrum is then given by
\begin{equation}
P_{\rm 2dF}(k) = \int d^3\veck' P_\zgal(|\veck-\veck'|) \hat{W}^2_{\rm 2dF}(k') \, ,
\end{equation}
where $\hat{W}$ is the normalized window function such that $\int d^3k\; W_{\rm 2dF}^2(k)=1$.

In Fig.~4, we show predicted redshift space galaxy power spectra convolved with the 2dF window function. 
As a comparison, we also plot the 2dFGRS power spectrum from Percival et al. (2001). Note that we have not attempted to model the
2dF data  by varying parameters of either the cosmological model or the halo description, 
but rather we show data to guide in understanding how non-linearities can affect current interpretations of the
2dF power spectrum;
A model fit to halo occupation number, based on the 2dF correlation function, can be found in Magliocchetti \& Porciani (2003).
In the middle and bottom panels of Fig.~4, we show the convolved redshift space power spectrum as a function of 
parameters in the halo description, mainly mass slope of the halo occupation number and the normalization of the virial
equation that determines galaxy motions within halos. For comparison, we also show the linear prediction,
again convolved with the 2dF window function. One should note the lack of oscillatory features in these power spectra when compared
to those in Fig.~3. This is due to the averaging over the 2dF window function and leads to a  reduction in
amplitude of small scale features (e.g., Elgaroy, Gramann \& Lahav 2002).

As shown in Fig.~4, the redshift space galaxy power spectrum is such that the transition to non-linearity comes with a reduction of
power, relative to the scaled linear power spectrum with a constant bias. This is opposite from the case associated with the
real space power spectrum where the on set of non-linearity is always associated with an increase in
power relative to the linear power spectrum scaled by a constant bias.
Another way to state this observations is that the bias factor for galaxies in redshift space, relative to linear clustering,
decreases from a constant value at large scales when non-linearities become important. As one moves to much small scales 
($ k > 10$ Mpc$^{-1}$), the non-linear redshift space power spectrum dominates over the linear description and the bias factor increases
back to a higher value (see, Fig.~3).  This is consistent with calculations of the redshift space bias by Seljak (2001). As shown in Fig.~4, we found the reduction in non-linear power to be present when certain parameters related to the halo model description is varied and we consider this to be a general feature of the redshift space
power spectrum.

In general, our calculations indicate that the linear assumption is only valid out to $\sim$ 0.1 Mpc$^{-1}$  both in the case of the real and redshift space
power spectra and proper modeling of these data beyond such a scale must include some accounting of non-linearities.
If measurements beyond 0.1 Mpc$^{-1}$ is used for cosmological purposes, with a constant bias factor, one can potentially
misidentify the drop in power due to non-linearities as due to a cosmological effect, for example, due to a massive neutrino since
massive neutrinos are expected to damp power at small scales.
In the next Section, we address this issue in detail using the Fisher matrix based approach to quantify how non-linearities
affect cosmological parameter determination and what improvements can be introduced to avoid potential biases.

\begin{figure*}
\centerline{\psfig{file=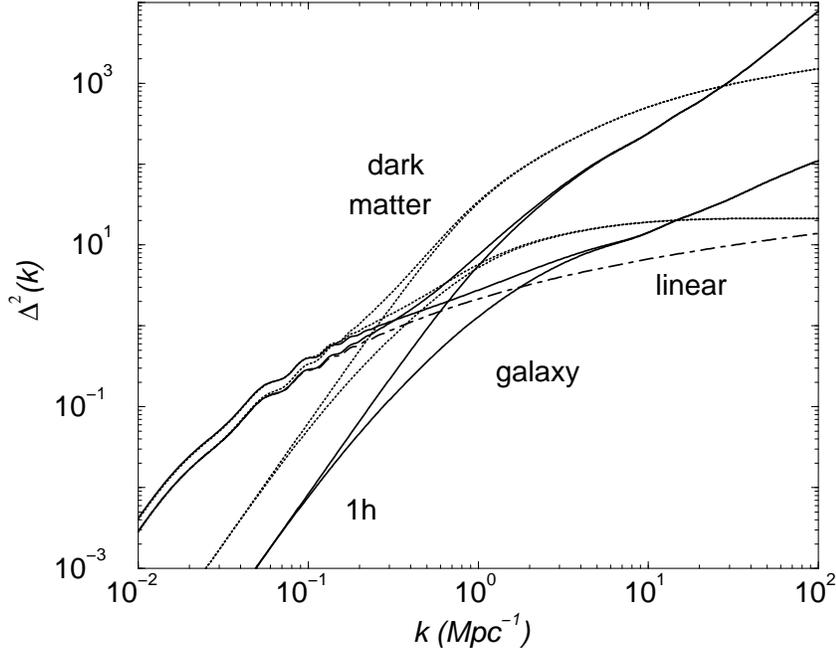,width=110mm,angle=-90}}
\caption{The power spectrum of dark matter (dotted lines) and galaxies (solid lines) 
in the redshift space. For comparison, we also show the real-space power spectra 
which is lower than the redshift space at large scales and higher at non-linear scales.
The increase in power at large scales is associated with bulk motions of halos while virial motions 
within halos lead to a suppression of power associated with the 1-halo term of the halo description.}
\label{fig:zspace}
\end{figure*}

\begin{figure}
\centerline{\psfig{file=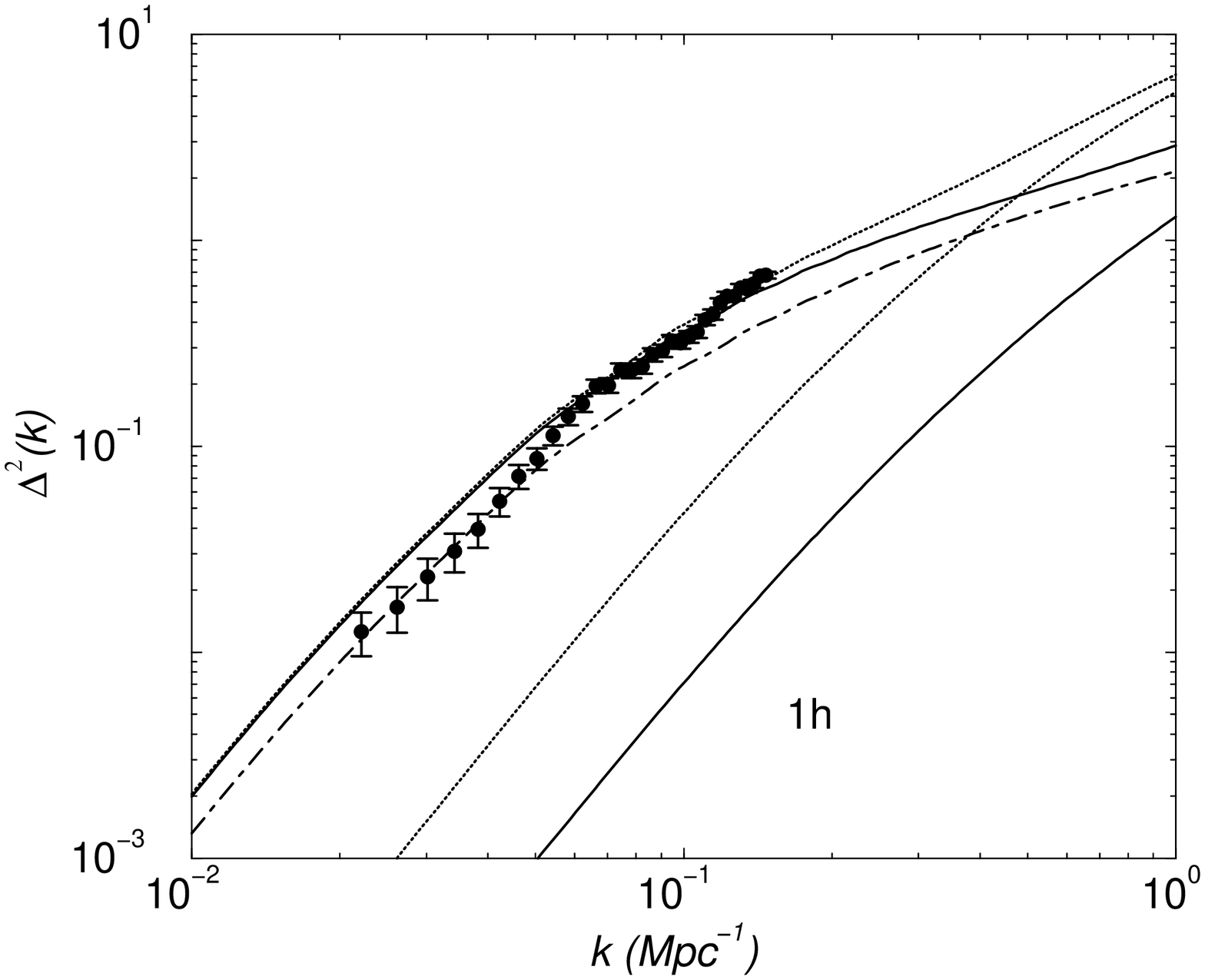,width=70mm,angle=-90}}
\centerline{\psfig{file=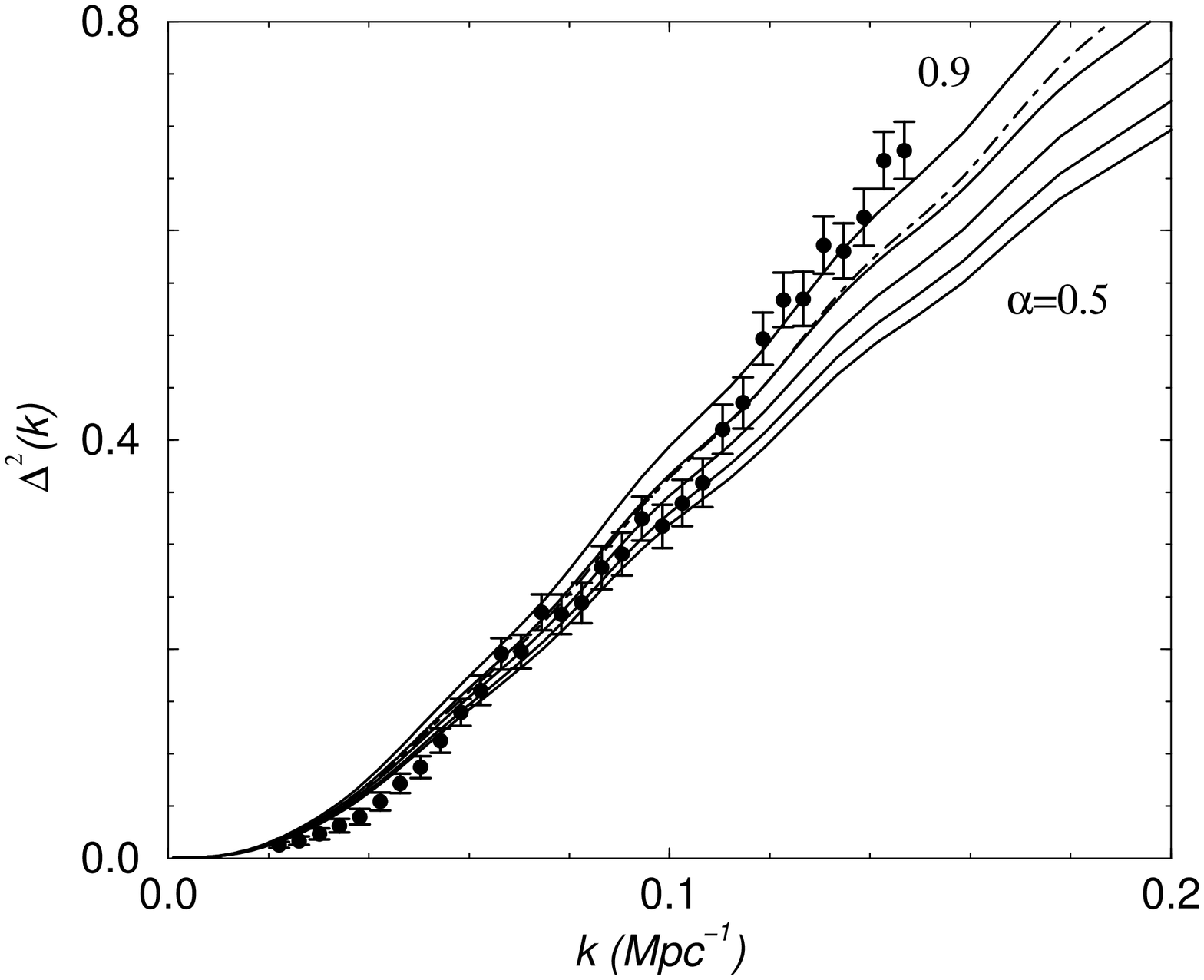,width=70mm,angle=-90}}
\centerline{\psfig{file=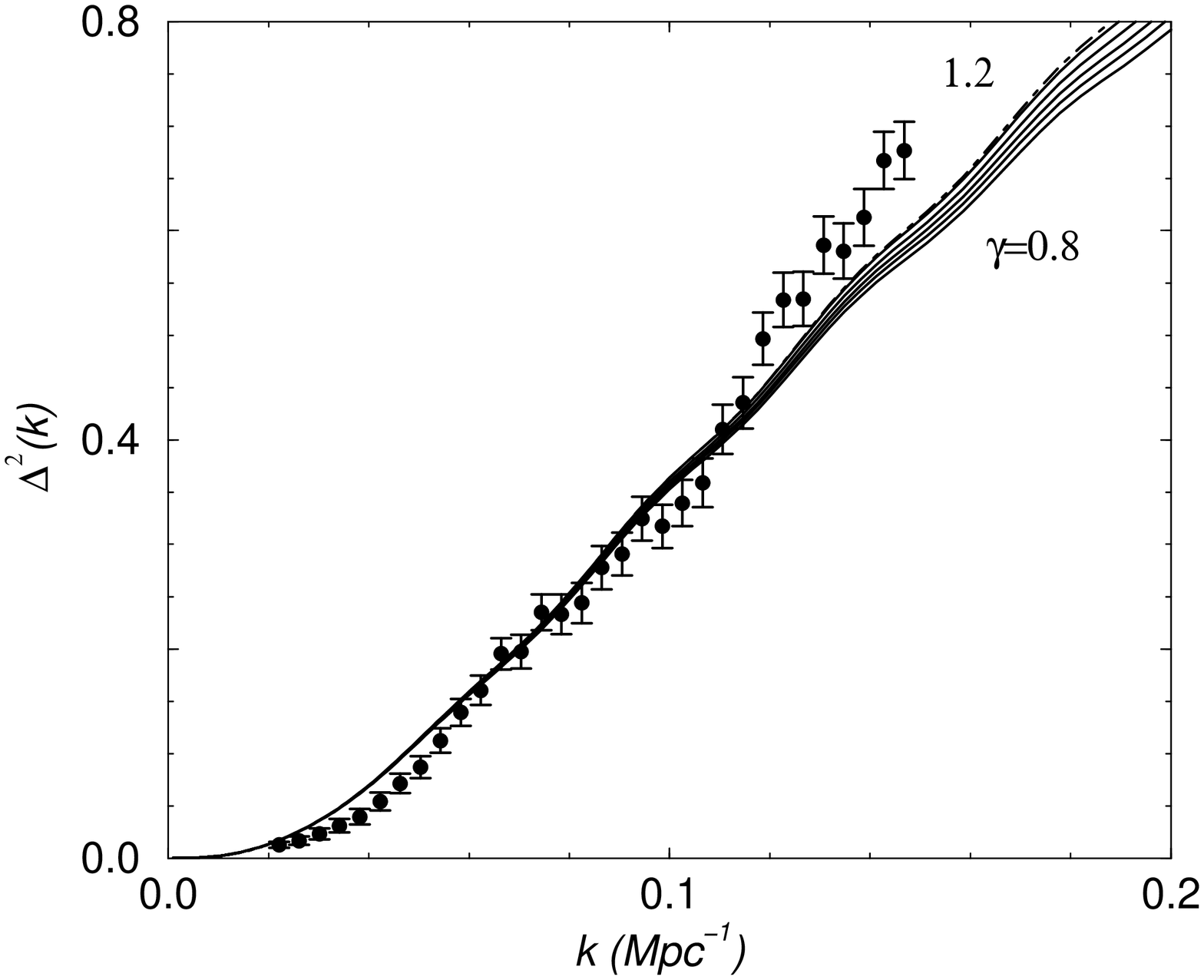,width=70mm,angle=-90}}
\caption{The redshift space power spectrum of the 2dFGRS (data points from Percival et al. 2001). The solid lines show the
redshift space galaxy power spectrum convolved with the 2dF window function. Note the lack of baryon oscillations features due to
smearing of modes over the finite width of the filter. For comparison, we also show the linear (dot-dashed line) and
the redshift space dark matter (dotted lines) power spectra, again convolved with the 2dF window function. The middle and bottom
panels are variations on the predicted redshift space galaxy power spectrum as a function of $\alpha$, the mass-slope of the
halo occupation number, and $\gamma$, the normalization of the  virial equation. In the middle panel, we assume $\gamma=1$ while
the bottom panel fixes $\alpha$ to be 0.8. In both case, $N_0$ is taken to be 0.7.}
\label{fig:2df}
\end{figure}

\begin{figure*}
\centerline{\psfig{file=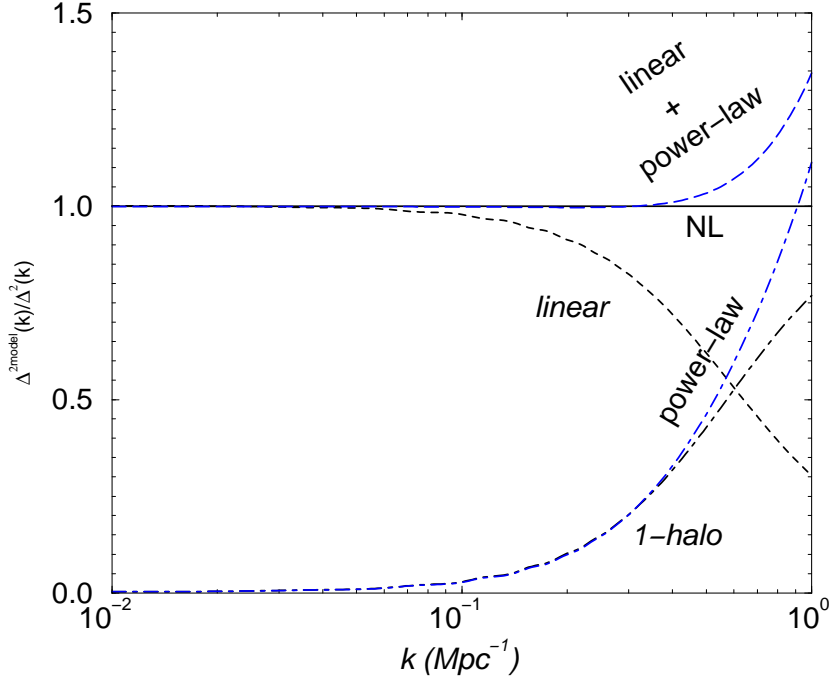,width=110mm,angle=-90}}
\caption{The ratio of modeled description of the galaxy power-spectrum to that of the total power spectrum as
expected under the halo model. The case illustrated here is same as the one shown in Fig.~2. With the
dotted line we show the linear power spectrum; as shown, the simple linear description with a constant bias
lead to an underestimate of true power at the level of 20\% when $k \sim 0.2$ to 0.3 Mpc$^{-1}$.
This can be compensated with the addition of a power-law contribution. The ratio related to the linear plus
a power-law is shown with a long-dashed line. The agreement is essentially exact out to 0.3 Mpc$^{-1}$, but
results  in an overestimate of power at deeply non-linear scales of 1 Mpc$^{-1}$ and at the level of 30\%.
For reference, with dot-dashed lines, we show the 1-halo contribution and the related power-law contribution
that attempts to model the non-linear aspect related to the power spectrum.}
\label{fig:ratio}
\end{figure*}

\begin{figure*}
\centerline{\psfig{file=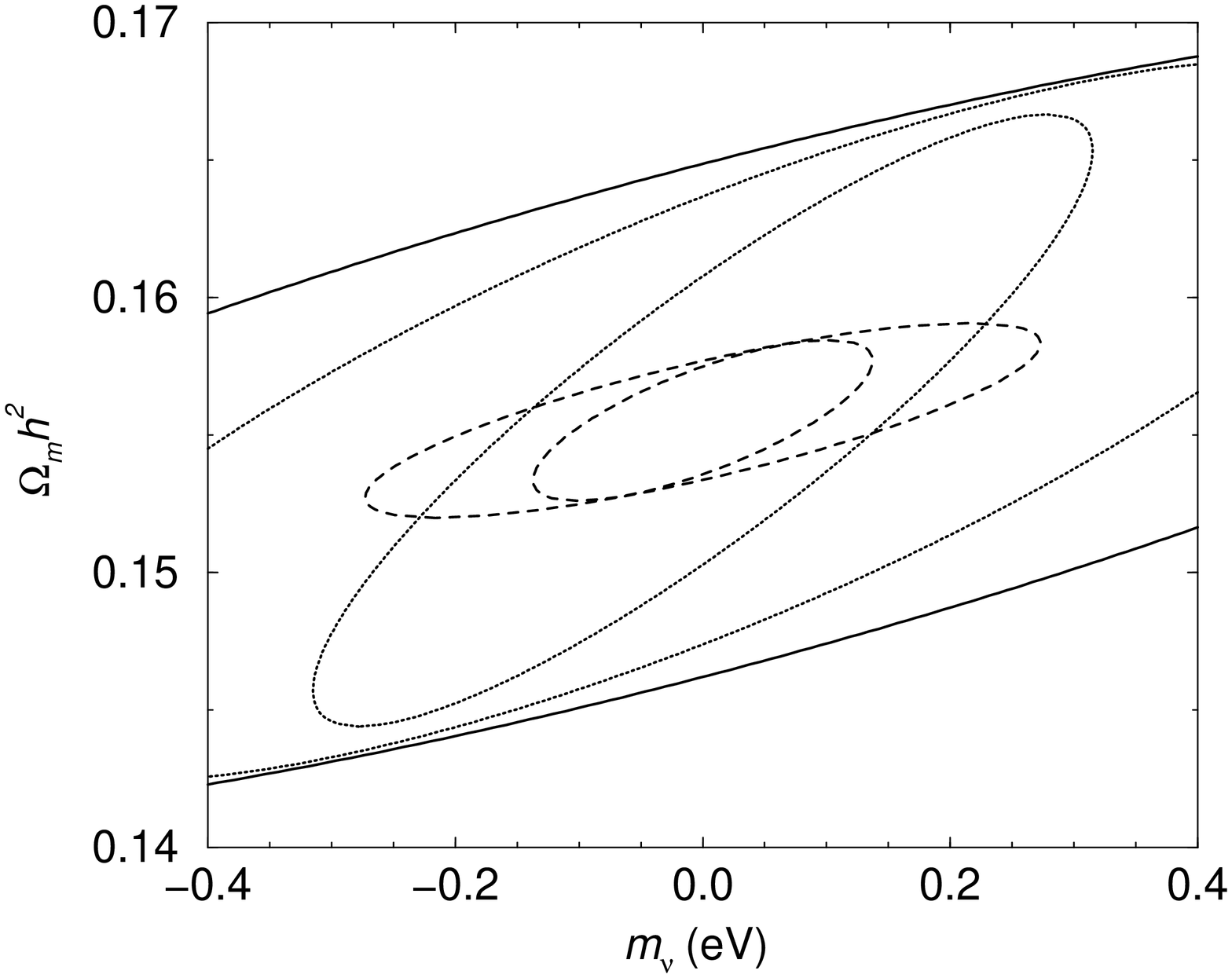,width=120mm,angle=-90}}
\caption{The 1$\sigma$ error ellipses for determination of $\Omega_mh^2$ and the neutrino mass. These errors follow 
from Tables 1 and 2. The solid line shows the constraint expected from the 3 year WMAP dataset (including polarization information).
The dotted lines show the expected improvement when galaxy clustering information is added with a cut off at $k < 0.1$ Mpc$^{-1}$
and using information out to 0.5 Mpc$^{-1}$ with an accounting for non-linearities (see text). The dashed lines show the same
situation but with Planck as the source of information related to CMB. The galaxy clustering survey considered here is
equivalent to that of the SDSS.}
\label{fig:neutrino}
\end{figure*}

\section{Non-linearities and Cosmology}

To quantify the extent to which cosmological parameters can be affected by assumptions in the literature
and to calculate expected errors on cosmological parameters and associated biases, we make use of
the Fisher information matrix:
\begin{equation}
{\bf F}_{ij} = -\left< \partial^2 \ln L \over \partial p_i \partial p_j
 \right>_{\bf x} \, ,
\label{eqn:likelihood}
\end{equation}
whose inverse provides the optimistic covariance matrix
for errors on the associated parameters (e.g., Tegmark et al. 1997).
In Eq.~\ref{eqn:likelihood}, $L$ is the likelihood of observing data set ${\bf x}$, in our case
the power spectrum, given the parameters $p_1 \ldots p_n$ that describe these data.
Following the Cram\'er-Rao inequality (Kendall \& Stuart 1969),
no unbiased method can measure the {\it i}th parameter
with standard deviation  less than $({\bf F}_{ii})^{-1/2}$ if all other parameters
are exactly known, and less than $[({\bf F^{-1}})_{ii}]^{1/2}$, from the inverse of the Fisher matrix,
 if other parameters are to be estimated from the  data as well.

For this discussion, we first focus on the binned real space  galaxy power-spectrum and return to the case involving the 
redshift space power
spectrum as part of the discussion. The binned power spectrum can be written as an
integral average over a shell in k-space centered around some wave number $k_i$ with width $\Delta k$:
\begin{equation}
\hat{P}_i = \frac{1}{V} \int_i \frac{d^3k}{V_i} P(k) \, ,
\end{equation} 
where $V_i = 4\pi k_i^2 \Delta k$ is the shell volume and $V$ is  the total survey volume.
The full covariance related to measurement of this binned power spectrum is \cite{Scoetal99,MeiWhi99,CooHu01}
\begin{eqnarray}
{\cal C}_{ij} &=& \langle \hat{P}_i \hat{P}_j \rangle - \langle \hat{P}_i \rangle \langle \hat{P}_j \rangle \nonumber \\
 &=& \frac{1}{V} \left[ \frac{(2\pi)^3}{V_i} 2 \hat{P'}_i^2 \delta_{ij} + T_{ij} \right] \, ,
\label{covariance}
\end{eqnarray}
where 
\begin{equation}
T_{ij} = \int_i \frac{d^3k}{V_i} \int_j \frac{d^3k}{V_j} T(\veck_i,-\veck_i,\veck_j,-\veck_j) \, ,
\end{equation}
is the non-Gaussian contribution to the covariance of the power spectrum. This non-Gaussian contribution depends on parallelogram
configurations of the trispectrum, $T(\veck_i,-\veck_i,\veck_j,-\veck_j)$, and
we make use of the 1-halo term of the halo approach \cite{CooShe02} 
to calculate it following Cooray \& Hu (2001). This approach is a valid description since in
linear scales the covariance is dominated by the Gaussian variance  while non-Gaussianities are only significant in
non-linear scales where the 1-halo term dominates. We can write the relevant contribution as
\begin{eqnarray}
&&  T_{ij} =   \int_i \frac{d^3k}{V_i} \int_j \frac{d^3k}{V_j} \nonumber \\
&&\int dm \, n(m) \,
  \frac{\left< N_\gal...(N_\gal-3)|m\right>}{\bar{n}_\gal^4}\,
  |u_\gal(k_i|m)|^2   |u_\gal(k_j|m)|^2 \,,\nonumber\\
\end{eqnarray}
where $\left< N_\gal...(N_\gal-3)|m\right>$ is the fourth moment of the halo occupation distribution. We follow 
our previous description related to the galaxy power spectrum and write $\left< N_\gal...N(\gal-3)|m\right>= \beta^2(m)[2\beta^2(m)-1][3\beta^2(m)-2]\left< N_\gal\right>^4$, where departures from the Poisson model, denoted with the function $\beta(m)$ in equation~\ref{binomial} 
is related to the adopted binomial distribution following 
Scoccimarro et al. (2001). In equation~\ref{covariance}, we allow for the shot-noise contribution to the power spectrum variance  by
redefining $\hat{P'}_i \rightarrow P_i + P^{\rm SN}$ such that the power spectrum is averaged over the shell with the shot noise contribution of $P^{\rm SN}=1/\bar{n}_g$, where  $\bar{n}_g = N_g/V$ is the number density of galaxies in the survey. In the Gaussian limit, the covariance only results in
a diagonal error matrix  with no correlations between band powers. This variance is given by the well known formula
\cite{PeaWes92,BlaGla03}
\begin{equation}
\sigma^2(k) = 2 \frac{(2\pi)^3}{V} \frac{1}{4\pi k^2 \Delta k} \left[P_\gal(k) + \frac{1}{\bar{n}_g}\right] \, .
\end{equation}

Using the full covariance,  the Fisher matrix for the galaxy power spectrum can be written as \cite{CooHu01}
\begin{equation}
{\bf F}_{\alpha \beta} = \sum_{ij} \frac{\partial \hat{P}_i}{\partial p_\alpha} \left[{\cal C}_{ij}\right]^{-1}\frac{\partial \hat{P}_j}{\partial p_\beta} \, ,
\end{equation}
where, for example,  $p_\alpha$ denotes a cosmological parameter of interest.
In addition to parameter errors, Fisher matrix approach also allows a useful technique to quantify the biases associated with
parameter estimates when the model description departs from the observations. For example, in the present case,
the galaxy power spectrum is usually modeled with a combination of the linear power spectrum and a scale-free bias out to a certain wavenumber, beyond which no information is used for cosmological purposes. We can quantify the extent to which this is valid by 
studying parameter biases such that if $p_\alpha = \bar{p}_\alpha+\delta p_\alpha$
where $\bar{p}_\alpha$ is the true value of the parameter $\alpha$ and $\delta p_\alpha$ is the resulting bias  due to the wrong
description of the power spectrum (in this case, linear power spectrum times constant bias), then
\begin{equation}
\delta p_\alpha = \sum_\beta F_{\alpha \beta}^{-1} \sum_{ij} \left[\bar{P}_i- \hat{P}_i\right] \left[{\cal C}_{ij}\right]^{-1}\frac{\partial \hat{P}_j}{\partial p_\beta} \, ,
\end{equation}
where $\hat{P}_i$ and $\bar{P}_i(k)$ represent the true and erroneous binned galaxy power spectra at wave number $k_i$, respectively (e.g., Huterer 2002).

To calculate errors, we assume a galaxy redshift survey consistent with the SDSS and take $V_{\rm SDSS}= 10^8$ h$^{-3}$ Mpc$^3$
and $N_g = 5 \times 10^5$ such that the galaxy density is $\sim 5 \times 10^{-3}$ h$^{3}$ Mpc$^{-3}$. 
For these assumed survey parameters, the non-Gaussian contribution to the covariance of the binned galaxy power spectrum is relatively
minor and results in a relative increase, when compared to the Gaussian covariance, by a factor of 1.4, at $k \sim$ 0.25 Mpc$^{-1}$
and a factor of 2 at $k \sim$ 0.5 Mpc$^{-1}$.   Note that if the number density of galaxies were to be higher, such that the shot-noise
is reduced at small scales, then the relative importance of  the non-Gaussian contribution to full covariance increases. 
The power spectrum is
assumed to be that modeled from the halo approach with $\alpha=0.8$ and $N_0=0.7$. We renormalized the power spectrum such that
the bias factor is unity at large scales. We then model the galaxy power spectrum through a combination of the linear
power spectrum and the constant bias, again assuming to be unity. The parameter errors and associated biases are tabulated in Table~1
for a $\Lambda$CDM cosmological model with a dark energy equation of state $w$ and allowing for massive neutrino contribution
to the mass density of the universe. We tabulated errors and biases as a function of the maximum $k$ value, $k_c$, to which galaxy
power spectrum analysis is considered to be linear. At typical large scales, with $k_c=0.1$ Mpc$^{-1}$, the biases are negligible and
we recover parameter errors usually quoted in the literature by the Fisher matrix based analyzes similar to the one performed
here \cite{Eisetal98,Wanetal99}.  In Table~1, we also tabulate errors expected from the galaxy power spectrum alone and the galaxy power spectrum combined with final WMAP data (middle rows) and with Planck data (bottom rows). The CMB information also includes polarization data in addition to temperature.

As one includes information out  to smaller scales, or large $k_c$ values, from the galaxy power spectrum,
 certain parameters, such as $\Omega_bh^2$ and
$m_\nu$, become better determined than out to an early cutoff, say at $k_c=0.1$ Mpc$^{-1}$. 
This can be understood by the fact that the small scale regime of the galaxy power 
spectrum captures information related to these parameters through additional oscillatory features and the damping of power by massive
neutrinos. This, however, comes at the expense of increasing biases in the parameter determination since in this regime,
non-linearities enter the galaxy power spectrum as the 1-halo term becomes important and starts to dominate
when $k > 0.6$ Mpc$^{-1}$ (see, Fig.~1).
Thus, the simple description based on the linear power spectrum and the scale-free bias breaks down when $k_c > 0.1$ Mpc$^{-1}$ though
there is significant amount of cosmological information beyond this cut off scale.
Note that in Table~1 we do not consider $k_c$ values above 0.5 since for the survey parameters we have considered,  the shot-noise
term associated with the finite number density of galaxies starts to become important. For future surveys that
attempt to measure an increased number of redshifts, it is clear that one gains further cosmological information 
by moving to smaller scales.

While cosmological parameter estimation from data below $\sim$ 0.1 Mpc$^{-1}$ is unaffected by non-linearities, it is useful to consider 
possibilities that can return information at wave numbers above this value. This is due to the fact that current surveys, both
2dFGRS and SDSS as well as previous surveys such as IRAS PCSZ, probed to much smaller scales then implied by this arbitrary cut off.
In the case of the real-space galaxy power spectrum, as shown in Fig.~1 (bottom panel), we consider the possibility that the onset of non-linearity can be modeled by
a power-law contribution when wave Numbers out to 1 Mpc$^{-1}$ is considered. 
In fact, for a good approximation, the 1-halo term can be modeled as a  power-law in the mildly non-linear regime with
\begin{eqnarray}
P_{1h}(k) = A_{\rm nl} k^{\alpha_{\rm nl}}   \; \; \;  k < k_1
\label{eqn:nl}
\end{eqnarray}
where $k_1$ is  the outer wave number to this approximation may be valid.
This description includes three new parameters involving the normalization of the non-linear
part of the power spectrum, $A_{\rm nl}$, the slope,
$\alpha_{\rm nl}$,  and the scale at which the power-law breaks, $k_1$.
Note that the power-law behavior remain the same when one considers different models of the halo
occupation number (Fig.~2).

The extent to which this approximation may be valid can be considered based on the halo model
description of the non-linear galaxy power spectrum with the addition of a power-law contribution.
While one can assume the halo model generated power spectrum as the expected non-linear power spectrum for
galaxies, in parameter estimation, one can consider the model fitting procedure based on
the approximation that the non-linear power spectrum is
$P(k) = b^2(k)P^{\rm lin}(k)+A_{\rm nl}k^{\alpha_{\rm nl}}$. 
The last two parameters account for
the non-linearity and, more importantly, the power law aspect of the 1-halo term. Since at large scales $u_\gal(k|m) \rightarrow 1$, one
expects the normalization to be given by
\begin{equation}
A_{\rm nl} = \int dm \, n(m) \,
                    \frac{\left< N_\gal(N_\gal-1)|m\right>}{\bar{n}_\gal^2}\, ,
\end{equation}
and the power-law $\alpha_{\rm nl} \sim 0$ (in Fig.~1, $\alpha_{\rm nl} \sim 3$ since we plot the logarithmic 
power spectrum, or $k^3 P(k)/2\pi^2$). The departure from this expected power-law captures the profile shape. Thus,
if these two parameters can be extracted from the data as well, one not only obtains information related to cosmology, but
some aspect related to how galaxies populate dark matter halos. The numerical value of $A_{\rm nl}$, for example, can
aid in constraining model independent measurements of the halo occupation number \cite{Coo02b}.

The extent to which this approximation may be valid is summarized in Fig.~5, for the same case shown in Fig.~2.
Here, we plot the ratio of the model power spectrum to that of the non-linear case as calculated by the halo model.
As already discussed, the description that $P(k) = b^2(k)P^{\rm lin}(k)$ breaks down rapidly such that
at a wavenumber of 0.1 Mpc$^{-1}$ one underestimates the true non-linear power by a few percent while at
0.3 Mpc$^{-1}$, this underestimate is at the level of 25\%. Instead of $P(k) = b^2(k)P^{\rm lin}(k)$, when
one includes the additional power-law such that $P(k) = b^2(k)P^{\rm lin}(k)+A_{\rm nl}k^{\alpha_{\rm nl}}$,
the difference is below a few percent out to a wavenumber of 0.3 Mpc$^{-1}$ suggesting that 
the transition is well modeled with the single power-law approximation. At a wavenumber of
0.5 Mpc$^{-1}$, the power-law assumption leads to an overestimate of power at a level of 5\%
and this overestimate is at the  level of 30\% when $k \sim 1$ Mpc$^{-1}$, deep in the non-linear regime.
While the single power-law assumption clearly breaks down at these small scales, the improvement over the
simple linear assumption is significant when $k \sim 0.3$ Mpc$^{-1}$.

While we have only considered a specific case with certain parameters for
the halo occupation number in Fig.~5, we expect the situation to be the
same with different descriptions. The extent to which one departs, however, is model
dependent as we have shown in Fig.~2, especially with respect to the
slope of the halo occupation number-halo mass relation (Fig.~2
top panel). The differences move the non-linear scale from 0.05 to 0.2
Mpc$^{-1}$ suggesting that the importance of the power-law varies over the
same scales for different samples of galaxies. The extent to which the
power-law approximation works will similarly vary and we note that
depending on $\alpha$, this can be between 0.3 and 0.5 Mpc$^{-1}$
while departures are at the level of 10\% to 50\% at 1 Mpc$^{-1}$; if the
power-law approximation is not included, the departures are substantial
($> 100$\%). Regardless of the exact variations, such as on the extent to
which the power-law assumption overestimate power when $k \sim$ 1
Mpc$^{-1}$, the power-law  assumption provides
a useful addition to the simple linear power spectrum description, with a
constant bias, in mildly non-linear
scales of order 0.2 to 0.4 Mpc$^{-1}$.

Note that the power-law assumption is independent of selection effects and biases related to observational aspects. This can be
understood simply as following. The observed total power spectrum is simply a convolution of the
real power spectrum time a window function that captures observational issues (Eq. 16).
Since the true power spectrum can be described better with a combination of the linear power spectrum and
a power-law correction, in the mildly non-linear scales, the observed power spectrum will remain
essentially described by the same combination with the filtering aspects related to observations applied.
Thus, regardless of how observations are done or the galaxy power spectrum is measured, we expect our description to 
be accurate to the level we can also describe the true underlying power spectrum with the same assumptions.

We can further understand the extent to which this model is valid by considering its ability to extract cosmological
parameters. In Table~2, we list expected errors on cosmological parameters in addition to these two new astrophysically oriented 
parameters . While cosmological parameter
estimates either remain same as in Table~1 or worsen slightly, we note there is a sharp reduction in biases associated with
parameter measurements when power spectrum information out to wave numbers of 0.5 Mpc$^{-1}$ is included. This is significant since
the assumed power-law behavior around the onset of non-linearity allows one to measure 
cosmology from previously ignored non-linear scales such that no biases are introduced. 
In Fig.~6, as an example, we show the measurement errors for two parameters ($\Omega_mh^2$ and $m_\nu$) by marginalizing over all other 
parameters. We show expected errors in combination with WMAP and Planck
and the improvement when information out to 0.5 Mpc$^{-1}$ is included with two additional parameters to be determined from data.
Including information out to 0.5 Mpc$^{-1}$ improves parameter errors, for example in the case of neutrino mass,
by roughly a factor of 2. 

So far, we have only considered the real space galaxy power spectrum. A similar situation exists for the redshift space power spectrum,
though, as discussed in Section~2.2, the onset of its non-linearity is associated with a decrease in power, relative to linear power 
spectrum times a constant bias, instead of an increase. To model this behavior,  one can introduce an additional damping term associated with
non-linear pairwise velocity to the usual formula involving the linear power spectrum \cite{Baletal96,HatCol98,Kanetal02}.
This leads to a modification to equation~\ref{eqn:deltaz} 
\begin{equation}
\delta_g^z(\veck) = \delta_g(\veck)\frac{[1+\beta\mu^2]}{\left[1+k^2\mu^2 \sigma^2/2\right]^{1/2}}\, ,
\end{equation}
such that the angular averaged redshift space power spectrum takes the form of
\begin{eqnarray}
&&P_{\zgal}(k) = \Big\{\frac{\left[\beta^2 k^2 \sigma^2 - 6 \beta^2 + 6 \beta k^2 \sigma^2\right]}{6 \left(k^2 \sigma^2/2\right)^2} \nonumber \\
&&+ \frac{\left(\beta - k^2 \sigma^2/2\right)^2 \arctan(\sqrt{k^2 \sigma^2/2})}{\left(k^2 \sigma^2/2\right)^{5/2}}\Big\} b_g^2 P^\lin(k) \, ,
\label{eqn:pzgal}
\end{eqnarray}
where $\beta=f(\Omega_m)/b_g$, as before,  and $b_g$ is the scale free large scale galaxy bias. 
When $\sigma \rightarrow 0$, this formula is equivalent to the standard formula used in
cosmological interpretations with the linear power spectrum scaled by the factor $(1+2/3\beta+1/5\beta^2)b^2_g$.
Note that such a correction only accounts for the increase in power at large scale associated with bulk motions.
At small scales, with wavenumbers $\sim$ 0.1 Mpc$^{-1}$  and larger, it is necessary to account for suppression of power
related to virial motions with dispersion $\sigma$, and we suggest the use of equation~\ref{eqn:pzgal}. 
In order to account for the 1-halo part of the non-linear redshift space power spectrum, especially for an
analysis of data beyond a wavenumber of $\sim$ 1 Mpc$^{-1}$, one can introduce again a power-law, following the suggested approach for the
non-linear real space power spectrum.  At mildly non-linear scales, the approach
suggested here leads to an additional parameter, $\sigma$, which can be simultaneously determined from data in
additional to cosmological estimates.
Currently, the well studied redshift space power spectrum comes from the 2dF survey and we plan to return to these issues as part of a combined modeling effort on non-linearities in this power spectrum based on cosmology as well as simplified corrections based on the halo approach.

\section{Summary}

The galaxy power spectrum is now a well known tool of precision cosmology.
Ongoing surveys such as 2dF and SDSS have already provided a wealth of information related to cosmological
parameters based on clustering of galaxies in the presumed linear regime, though
measurements span down to much smaller scales than used for cosmological purposes.
We have shown here that there is considerable amount on information in the regime where clustering transits
from linear to non-linear. The extraction of this information, however, cannot be done alone with the usual
approach involving the linear power spectrum and a constant scale-free bias. We have investigated the
non-linear behavior of the galaxy power spectrum in both real and redshift space based on the halo approach
and have suggested that the on set of
non-linearities can be modeled as a simple power-law. The normalization and the power-law slope provide certain information related to halo occupation description in the halo models of galaxy clustering.
We have shown that significant improvements can be made when non-linear effects are taken into account 
through this power-law contribution  with two additional parameters to be determined from the data simultaneously
with cosmological parameters. The suggested approach can easily be implemented in current cosmological studies of
galaxy clustering and to extract most information from galaxy clustering data to date. 

\section*{Acknowledgments}
 
We acknowledge useful discussions with O. Lahav on the 2dFGRS power spectrum. This work is supported by the Sherman Fairchild foundation and
DOE DE-FG 03-92-ER40701. AC thanks Kvali Institute for Theoretical Physics (supported by NSF PHY99-07949) for hospitality 
where this work was initiated.

\end{document}